\def\year{2020}\relax
\documentclass[letterpaper]{article} % DO NOT CHANGE THIS
\usepackage{aaai20}  % DO NOT CHANGE THIS
\usepackage{times}  % DO NOT CHANGE THIS
\usepackage{helvet} % DO NOT CHANGE THIS
\usepackage{courier}  % DO NOT CHANGE THIS
\usepackage[hyphens]{url}  % DO NOT CHANGE THIS
\usepackage{graphicx} % DO NOT CHANGE THIS
\usepackage{graphicx}  % DO NOT CHANGE THIS
\usepackage{amssymb}
\usepackage{algorithm}
\usepackage{algpseudocode}
\usepackage{mathtools}
\usepackage{multirow}
\usepackage{array}
\usepackage{subfig}

\makeatletter
\newcommand{\algmargin}{\the\ALG@thistlm}
\makeatother
\algnewcommand{\parState}[1]{\State%
    \parbox[t]{\dimexpr\linewidth-\algmargin}{\strut\hangindent=\algorithmicindent \hangafter=1 #1\strut}}

\title{Heterogeneous Multi-Agent Reinforcement Learning for Unknown Environment Mapping}
\author{Ceyer Wakilpoor\textsuperscript{\rm 1}, Patrick J. Martin\textsuperscript{\rm 2}, Carrie Rebhuhn\textsuperscript{\rm 1}, Amanda Vu\textsuperscript{\rm 1}\\
\textsuperscript{\rm 1}The MITRE Corporation, \textsuperscript{\rm 2}Virginia Commonwealth University\\
cwakilpoor@mitre.org, martinp@vcu.edu, crebhuhn@mitre.org, amandavu@mitre.org}

\begin{document}

\maketitle

\begin{abstract}
Reinforcement learning in heterogeneous multi-agent scenarios is important for real-world applications but presents challenges beyond those seen in homogeneous settings and simple benchmarks. 
In this work, we present an actor-critic algorithm that allows a team of heterogeneous agents to learn decentralized control policies for covering an unknown environment. 
This task is of interest to national security and emergency response organizations that would like to enhance situational awareness in hazardous areas by deploying teams of unmanned aerial vehicles.
To solve this multi-agent coverage path planning problem in unknown environments, we augment a multi-agent actor-critic architecture with a new state encoding structure and triplet learning loss to support heterogeneous agent learning.
We developed a simulation environment that includes real-world environmental factors such as turbulence, delayed communication, and agent loss, to train teams of agents as well as probe their robustness and flexibility to such disturbances.
\end{abstract}

\section{Introduction}\label{sec:intro}

Over the past ten years, unmanned aerial vehicles (UAVs) capabilities have advanced considerably thanks to open source software and hardware as well as new perception and estimation algorithms. U.S. national security and emergency response organizations would like to deploy these systems in hazardous, tactical situations. Although these UAVs are readily available, their use requires trained pilots to operate them safely in these hazardous zones. Many research efforts across defense and civilian agencies are developing new software and hardware technologies that enable more autonomous capabilities, but the state-of-the-art solutions still require operator babysitting \cite{ras:2017}.

One critical government use case for UAVs is the rapid update of tactical maps in hazardous regions.
This use case is well suited for multiple UAVs, since their individual endurance may not last the entire mission length. 
Academic research in this field has largely focused on developing more efficient multi-vehicle path planning algorithms. 
Although these solutions have optimality guarantees, they require \emph{known}, \emph{static environments}.
Obstacles (e.g. rubble, toppled buildings) and adversarial agents make these solutions brittle when deployed in dynamic scenarios such as battlefield reconnaissance or disaster recovery.

Reinforcement learning (RL) has recently been applied to challenging domains ranging from game playing \cite{mnih:2015,silver:2016} to robotic manipulation and navigation tasks \cite{levine:2016,zhu:2017}. 
The RL paradigm shifts the computational burden to offline training, with agents learning from experience during execution. 
In the multi-agent RL case, the real-time execution of agent policies can lead to complex, cooperative behaviors. 
Multi-agent RL systems have been applied to the StarCraft unit micromanagement task \cite{vinyals:2017}, as well as other simple scenarios such as landmark coverage and cooperative `go-to-point' tasks.
These current methods focus on homogeneous team compositions (a more manageable setting that allows for parameter sharing between actors) and simpler tasks that are well-defined with little environment stochasticity, such as games.
Many challenges remain on how to develop multi-agent RL methods to solve more complex, real-world tasks involving heterogeneous team compositions and robustness to environmental effects. 

In this work, we propose a multi-agent RL approach to covering unknown environments that is flexible to teams with heterogeneous sensor payload compositions. Our approach, called embedded multi-agent actor-critic (EMAC), is flexible to dynamic environment elements not present in traditional coverage path planning algorithms. We adopt a multi-agent actor-critic framework and introduce a new encoder network that ingests variable-length agent observations and outputs fixed-length observation encodings. The fixed-length observation embeddings allow for teams of agents with heterogeneous sensor payloads. In cooperative settings, the fixed-length encodings have the added benefit of allowing for parameter sharing among actors, similar to the homogeneous team case. The observation encodings themselves are subjected to a triplet learning loss where multiple, simultaneous viewpoints of the same global observation are attracted in the embedding space, while being repelled from temporal neighbors which are often visually similar but functionally different \cite{sermanet:2017}.
	
Our results show that EMAC outperforms traditional multi-vehicle path planning algorithms as well as independent RL baselines. Overall, EMAC's ability to dynamically react to the environment and adjust according to the actions of its team members allows for more efficient, cooperative behaviors to emerge. 
This is partially due to our novel pixel-based state representation with near-field and far-field views, which  outperforms traditional coordinate-based state representations. 
We probe the robustness of our approach by introducing real-world environmental effects such as communication delays, agent drop out, and wind turbulence to the simulation. 
We evaluate the scalability of EMAC to increasing environment size and numbers of agents, finding that EMAC scales well with environment and team sizes.
Finally, we assess the heterogeneous capabilities of EMAC, finding that our approach is flexible and robust to a variety of team compositions.

\section{Related Work}\label{sec:relatedwork}

Multi-agent coverage path planning (CPP) over unknown terrain is a task that applies to national security and emergency response scenarios where rapid tactical map updates in hazardous regions would boost situational awareness. 
In the CPP problem, agents require paths that cover an environment of interest while minimizing overlap and avoiding obstacles \cite{galceran:2013}.

Typically, CPP algorithms assume a \emph{fixed}, \emph{known environment}. 
These algorithms decompose and encode the environment into a data structure representation, such as a graph. 
The costs of model motion within the data structure are calculated and an optimal workload assignment problem is solved across available agents\cite{barrientos:2011,modares:2017,maza:2007}. 
Since these approaches require full environment knowledge, it makes them ill-suited to CPP in unknown terrain, where they may encounter dynamic obstacles, environmental variations, or adversarial agents. 
An alternative approach is to shift the planning algorithm computation to a simulation environment and apply reinforcement learning. 
To the best of our knowledge, our paper is the first application of multi-agent RL to the CPP problem and provides a novel approach to heterogeneous agent learning. 

Multi-agent RL is a long-studied problem that in the past few years has started moving from tabular algorithms to deep learning methods capable of tackling high-dimensional state and action spaces \cite{tampuu:2017,foerster:2018,mnih:2016}. 
One approach to learning multi-agent RL policies is to directly learn policies and value functions in a fully decentralized manner. Independent Q-Learning (IQL) trains independent action-value functions for each agent using Q-Learning \cite{tampuu:2017}. Similarly, Independent Actor-Critic (IAC) has each agent learn independently with its own actor and critic \cite{foerster:2018}. These decentralized approaches suffer from some limitations, such as non-stationarity of the environment from the perspective of individual agents. 

Recent approaches have utilized a variation of actor-critic frameworks \cite{mnih:2016} to enable centralized training with decentralized execution, circumventing the environment non-stationarity problem of independent-learner approaches. Most notable from the actor-critic approaches are COMA \cite{foerster:2018} and MADDPG \cite{lowe:2017}. COMA uses a centralized critic to estimate the action-value function and decentralized actors to optimize the agents' policies. 
To address the challenges of multi-agent credit assignment, COMA introduces a counterfactual base line that marginalizes a single agent’s action while keeping the other agent actions fixed. Similarly, MADDPG also uses decentralized actors to optimize agent policies, but employs separate centralized critics for estimating the action-value function. MADDPG is applicable to cooperative and competitive environments.

To date, most work in multi-agent RL has focused on homogeneous team compositions. 
Team homogeneity allows for parameter sharing among actors as well as simpler network architectures, which leads to faster and more stable training.
However, in real-world applications, it is likely that a multi-agent team will have a heterogeneous composition. 
Agents will need to leverage their unique abilities and rely on other agents' specializations to cooperate and find effective policies. 
As world dynamics become more complex and the agents have less shared behaviors (i.e., different actions, domain knowledge, goals), learning becomes more difficult. Consequently, heterogeneous learning processes are significantly slower and harder to optimize than their homogeneous counterparts. The StarCraft unit micromanagement baselines are an example of a multi-agent RL baseline which tests on a mixture of both homogeneous and heterogeneous scenarios. However, it is worth noting that while the agents on a team may have different specialties (for example, Stalkers versus Zealots), the observation and action spaces for both unit types are the same.

\section{Background}\label{sec:background}

Here, we describe the baselines for the unknown environment coverage task. 
We implement a non-RL baseline derived from CPP literature. 
In addition, we use two RL baselines, deep Q-learning and actor-critic, which will be extended to multi-agent settings as independent Q-learning (IQL) and independent actor critic (IAC). 

\subsection{Notation}

We consider unknown environment coverage as a fully cooperative multi-agent task. 
This problem is modeled as a decentralized partially observable Markov decision process (Dec-POMDP) where we define $s_t$ as the true state of the environment at time $t$.
We assume that there are $n$ agents, indexed by $i={1, ..., n}$, and each agent takes an action from its action set, $U^i$.
At each time step, agent $a^i$ will choose an action $u_t^i\in U^i$ according to their partial observation of the environment, $o_t^i$. 
The teams' joint action induces a transition in the environment and a corresponding reward $r(s_t, u_t)$, where $u_t = [u_t^1, \cdots,u_t^n]$. 

The resulting action-observation history of each agent can be used to learn a stochastic policy $\pi$. 
The joint policy induces a value function $V^{\pi}(s_{t}) = {\mathbb E}_{\pi} [R_{t} | s_{t}]$, or an action-value function $Q^{\pi}(s_{t}, u_{t}) = {\mathbb E}_{\pi} [R_{t} | s_{t},u_{t}]$, where $R_{t} = \Sigma^{\infty}_{t=0} \gamma^{t} r_{t+1}$ is the discounted return and $\gamma \in [0, 1)$ is a discount factor. 
We compute the advantage function with $A^{\pi}(s_{t},u_{t})=Q^{\pi}(s_{t},u_{t}) - V^{\pi}(s_{t})$.

\subsection{Non-RL Approach (NRL)}

The non-RL baseline decomposes the environment into a graph and assigns areas based on optimal workload. 
In our implementation, we use Voronoi decomposition to partition the environment into Voronoi cells. 
Locations near cell edges are placed in a graph structure, and their visitation order is determined by a greedy heuristic based on the number of unvisited surrounding cells. 
A spiral pattern covers the nodes and the center of the cell. 
Agents avoid obstacles using breadth first search to find free cells around the obstacles. 
In the unknown environment coverage task, the NRL approach has advantages over RL approaches since it performs decomposition in a fully observed world; however the agents will not be able to react to dynamic obstacles.

\subsection{Independent Q-Learning (IQL)}

Deep Q-learning techniques \cite{mnih:2015} aim to solve for the action-value function $Q^{*}(s_t,u_t)$ corresponding to the optimal policy. $Q^{*}$ is represented as a deep neural network parameterized by $\theta$ and found by minimizing the loss $\mathcal{L}(\theta)=  {\mathbb E} [(Q(s_t,u_t;\theta)-y)^{2}]$ where $y=r(s_t,u_t)+\gamma\max_{u'}[Q(s'_t,u'_t;\theta')]$ and $\theta^{-}$ are the parameters of a target network that is periodically copied from $\theta$.

Q-learning can be directly applied to multi-agent settings by having each agent $a^i$ learn independently optimal $Q_{a^i}$, a method known as independent Q-learning \cite{tampuu:2017}. 
Although this approach does not address non-stationarity issues arising from the training process, in practice IQL serves as a benchmark for cooperative tasks.

\subsection{Independent Actor-Critic (IAC)}

Actor-critic methods \cite{mnih:2016} jointly optimize for the actor and the critic by using the gradient estimated by the critic to train the actor. 
The critic, i.e. the action-value function $Q$ or value function $V$, is represented by a deep neural network parameterized by $\phi$ and found by minimizing the loss (in the action-value case): $\mathcal{L}(\phi) = {\mathbb E} [A(s_{t},u_{t})]^{2}$. The actor, i.e. the policy function $\pi$, is represented by a deep neural network parameterized by $\theta$ and found by minimizing the loss $\mathcal{L}(\theta) = {\mathbb E}[\log\pi(u_{t}|s_{t};\theta)A(s_{t},u_{t})]$. Independent Actor-Critic (IAC) is a simple extension of the actor-critic approach to multiple agents where each agent learns independently, each with its own actor and critic, and only from its own action-observation history \cite{foerster:2018}. 

\section{Technical Approach}\label{sec:techapproach}

\begin{figure}[t]
    \includegraphics{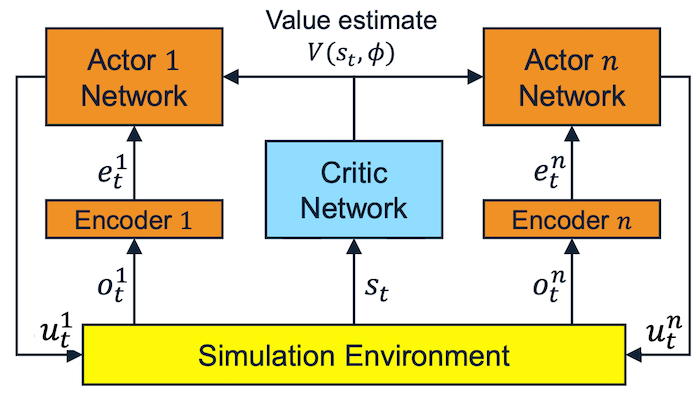}
    \centering
    \caption{This figure illustrates the embedded multi-agent actor-critic (EMAC) architecture. During training, EMAC utilizes a shared centralized critic to estimate the value function, $V(s_t,\phi)$. To enable heterogeneous teams, each agent encodes observations using a fully connected layer. At execution time, the critic is removed and agents execute their policies individually. }
    \label{fig:arch}
    \end{figure}

This section describes our embedded multi-agent actor-critic (EMAC) approach to solving the CPP problem. Figure \ref{fig:arch} illustrates this multi-agent actor-critic approach that uses embedding networks to support heterogeneous multi-agent learning on cooperative tasks.

\subsection{Embedding Network}

The embedding network is a key addition to the actor-critic framework that allows EMAC to function with heterogeneous team compositions. 
The embedding network is a fully connected layer that is assigned to each agent in the team. 
Variable length observations are fed to the embedding network and encoded into fixed length feature vectors. 
Since encoded observations are now a fixed length, actor parameters are shared during training, as in the homogeneous case. Parameter sharing among actor networks accelerates and stabilizes learning.
%% PJM: I took out the memory claim as we do not have a benchmark number to prove it and reviewer 4 whined about it a bit.
%% Probably best to leave it out.
These encoder networks enable heterogeneous sensing capabilities while retaining the training benefits of homogeneous teams. 

\begin{figure}[t]
	\centering
   	\includegraphics{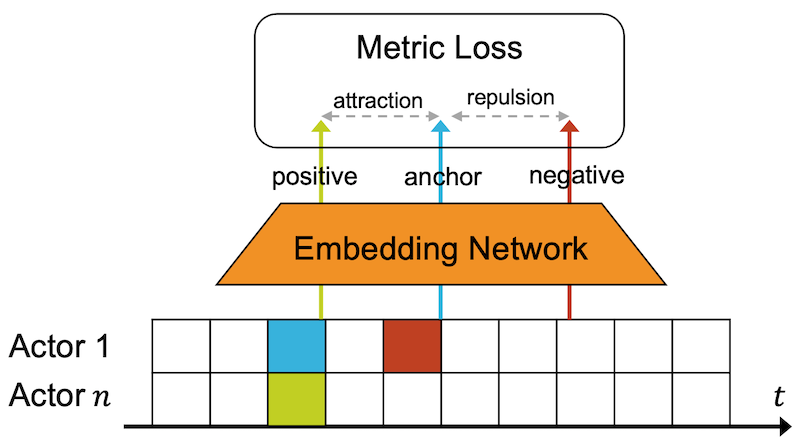}
	\caption{Triplet loss. This loss is used to encourage consistent embedding representations across the agent encoder networks. The anchor (blue) is associated with positive samples (green) which come from different actors at the same time step, and negative samples (red) which are taken from different times within the same actor sequence.}
	\label{fig:tripletloss}
    \end{figure}

In order to encourage consistency across the different embedding networks, the observation encodings are subjected to a triplet loss inspired by the multi-view metric learning loss used in self-supervised imitation learning \cite{sermanet:2017}. The idea behind the triplet loss is to encourage observations coming from the same time but different viewpoints (i.e., different agents) to be pulled together, while pushing apart visually similar observations from temporal neighbors. 
This time-contrastive loss provides a strong training signal to the embedding network to learn useful embedding representations that can explain environmental changes over time. More formally, given an observation $o_t^i$, the observation encoding can be represented as $f(o_t) \in \mathbb{R}^{d}$, where $o_t=[o_t^1,\cdots,o_t^n]$. 
We set the observation history of an individual agent $f(o_t^i)$ as the anchor and the co-occurring observation history of one of its teammates $f(o_t^j), j \neq i$ as the positive sample. The negative sample $f(o_{\sim t}^i)$ consists of temporal neighbors of the anchor that exceed a minimum time buffer around the anchor. 
Thus, the system learns an embedding $f$ such that $\parallel f(o_t^i)-f(o_t^j)\parallel_{2}^{2}+ \alpha<\parallel f(o_t^i)-f(o_{\sim t}^i)\parallel_{2}^{2}$, where $\alpha$ is a margin that is enforced between positive and negative pairs. 
A schematic illustrating the soft-margin triplet loss can be seen in Figure \ref{fig:tripletloss}.

\subsection{Multi-Agent Advantage Actor Critic}

EMAC builds off of the multi-agent advantage actor-critic framework \cite{foerster:2018} that facilitates centralized policy training and decentralized policy execution. 
A critic is shared among all the agents and acts as the centralized value function during training.
Its input information is the global state of the environment, $s_t$, and it outputs the estimated state values $V(s_{t},\phi)$. The critic parameters $\phi$ are updated by mini-batch gradient descent, minimizing the following loss: $\mathcal{L}(\phi)= \mathbb{E}[A(s_{t},u_{t} )^{2}]$, which functionally can be rewritten as 

\begin{equation} \label{eqn:critic}
	\mathcal{L}(\phi)= \mathbb{E}[(R_{t}-V(s_{t},\phi))^{2} ]
	\end{equation}
since $R(t)$ is an estimate of $Q^{\pi}(s_{t}, u_{t})$ and $V(s_{t},\phi)$ is an estimate of $V^{\pi}(s_{t})$.

The actor network represents the agent policy function. 
A fixed-length encoded observation, $f(o_t^i)$, is fed into this network, so it is shareable among heterogeneous agents. 
The actor loss is defined as $\mathcal{L}(\theta)=\mathbb{E}[\log\pi(u_{t}|s_{t})A(s_{t},u_{t})]$, where $A(s_{t},u_{t})$ is the same advantage computation used for the critic. 
We introduce an entropy term to encourage exploration and avoid converging to non-optimal policies \cite{harnooja:2017}: $\mathcal{L}_{H}=\pi(u_{t},s_{t})\log\pi(u_{t},s_{t})$. 
This term leads to the following overall loss for the actor network:

\begin{equation} \label{eqn:actor}
\begin{aligned}
	\mathcal{L}(\theta)=\mathbb{E}[\log\pi(u_{t}|s_{t})A(s_{t},u_{t})] \\
						+ \ \pi(u_{t},s_{t})\log\pi(u_{t},s_{t})
	\end{aligned}
\end{equation}
During offline learning, the actor and critic are trained jointly. 
During execution, the critic is removed and the agents select actions in a decentralized manner. 

\subsection{EMAC Algorithm}

EMAC is trained using parallel rollout threads, with each thread initialized with its own environment instance \cite{mnih:2016}. At the end of each episode, mini-batches are formed across the parallel environments and used to make Monte Carlo (MC) updates for the encoder, actor, and critic networks. The environments are reset before beginning another training episode. Empirically, we found that training with parallel rollout threads had a stabilizing effect on the learning process. 

\begin{algorithm}
  \caption{Training procedure for EMAC}
  \begin{algorithmic}[1]
  	\State Initialize $E$ parallel environments with $n$ agents
  	\For{$episode=1...max\_episodes$}
  		\parState{Reset environments and get initial observations $o^{i,e}_{t}$ for each agent $i$ in each environment $e\in E$}
  		\State Reset episode buffer $D$
  		\For{$t=1...max\_steps$}
  			\parState{Apply embedding network: $f(o^{i,e}_{t})$}
  			\parState{Step agents: $u_t^{i,e} \sim \pi(\cdot \mid f(o^{i,e}_{t}))$}
  			\parState{Send actions to all environments to get $\hat{o}^{i,e}$, $r^{i,e}$}
  			\parState{Store($o^{i,e},\hat{o}^{i,e},r^{i,e}$) in $D$}
  			\If{all episodes ended}
  				\parState{Unpack minibatch $(o, \hat{o}, r) \leftarrow D$}
  				\parState{Update critic according to Eqn (\ref{eqn:critic})}
  				\parState{Update actor according to Eqn  (\ref{eqn:actor})}
  				\parState{Update encoder according to triplet loss}
  				\parState{\textbf{break}}
  			\EndIf
  		\EndFor
  	\EndFor
  \end{algorithmic}
\end{algorithm}

\section{Unknown Environment Coverage Task}\label{sec:unknownenvcoverage}

This work investigates a cooperative multi-agent task: unknown environment coverage. 
This task is relevant for government agencies that would like to use autonomous systems for tactical map construction in hazardous areas.
In our simulated scenario, a team of agents are randomly placed in a 2D grid world, and tasked with covering the entire grid using their imaging sensors.
Random locations in the grid world contain terrain obstacles that cannot be traversed by an agent. If an agent takes an action such that it collides with obstacles or other agents, it is considered inactive and will not take actions in the environment. 
Each agent occupies one cell of the grid and has an imaging sensor that covers a $k \times k$ subset of the environment. 
The episode ends when either all the cells in the grid have been covered by the agents' imaging sensors, or when the maximum episode timeout has been reached. 

The simulation environment facilitates exploration of heterogeneous agent teaming, as well as configurable environmental effects that would likely affect agent task execution. 
These features provide a path toward more realistic RL training scenarios that would be of interest to multiple government organizations.

\subsection{Agent Information}

\begin{figure}[t]
	\centering
   	\includegraphics{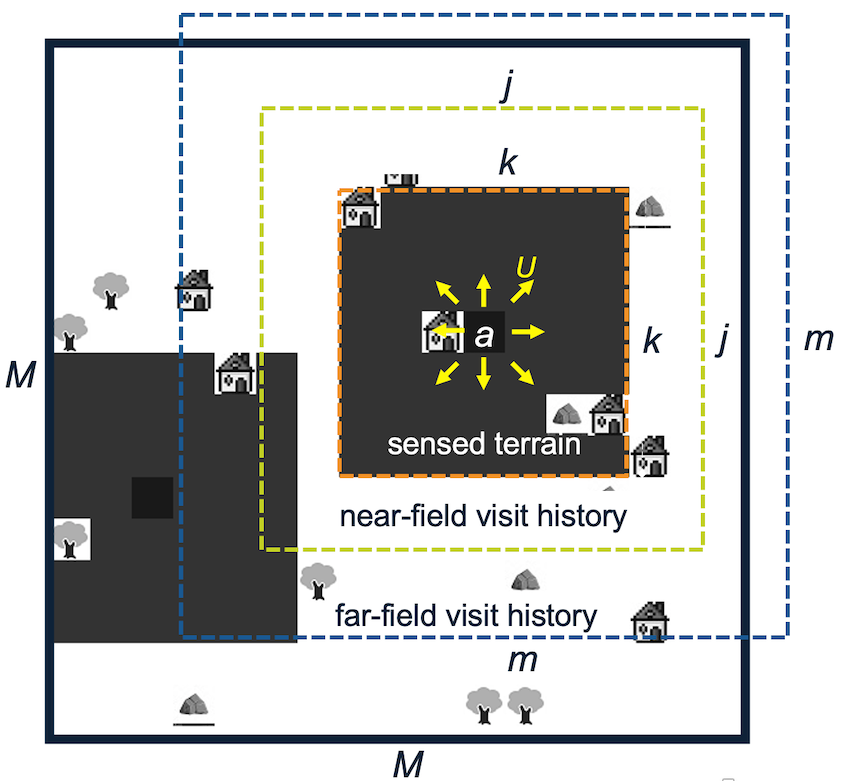}
	\caption{The states and actions of an agent at the first timestep of the unknown environment coverage task. Agents (black) are placed in an environment with randomly generated obstacles (mountain, house, tree icons). Each agent is equipped with an environment sensor with a $k \times k$ field of view (grey). With no previous knowledge of the environment, the agents must map the entire grid (i.e. view all squares in the grid with their imaging sensors) as quickly as possible.}
	\label{fig:agentinfo}
    \end{figure}

In Figure \ref{fig:agentinfo}, we illustrate the components of the agent action and observation space.
The state space is represented as an $\mathbb{R}^{M \times M}$ map, with terrain and visitation information encoded in each cell. 
Each agents' action set, $U^i$, has eight cardinal direction actions as well as a `no-move' action. 
During training, actions that result in collisions are converted to a no-move action. 

An agent observation is a concatenated tuple of four different inputs: a $k \times k$ sensed terrain map, a $j \times j$ near-field visit history map, an $m \times m$ adaptively-pooled  far-field visit history map (where $m \leq 2M$ and is padded if off-map), and the agents' action at the last time step. 
Each input is scaled to be between 0.0 and 1.0. 
Empirically, we found that our egocentric pixel-based representation converged to better policies in comparison to coordinate-based state representations traditionally used in navigation tasks. 
The combination of near-field high-resolution and far-field low-resolution pixel-based views serve as a rich state-representation space that is well-suited to navigation tasks in cluttered terrain. 

The reward formulation consists of both team-based rewards and individual rewards. 
The rewards include: (1) a team-based terminal reward given after successfully covering the grid, (2) a team-based progress reward  based on the fraction of unseen cells uncovered during a time step, (3) an individual discovery reward given if an agent uncovered a previously unseen cell with their imaging sensor, (4) an individual visitation penalty if the agent did not uncover any previously unseen cells, and (5) an individual collision penalty if the agent collided with terrain or went out of bounds.

\subsection{Environmental Factors}\label{sec:env_factors}

Our simulation implements several environmental effects that simulate the stochastic nature of real-world settings.
The environmental factors probe the robustness of RL agents in more mission-relevant scenarios than what other popular benchmarks provide.

\subsubsection{Communication delay}

Agents communicate their position in the environment to their team mates. 
When enabling communication delay, messages passed between team members may be delayed by $n$ time steps. 
This effect simulates network degradation over long distances or corrupted network communications. 

\subsubsection{Agent dropout}

This effect causes an agent to drop with probability, $p$, until a minimum number of agents are left in the environment. 
Dropped agents will no longer take actions in their environment. 
Agent drop out simulates agent failure due to adversarial effects or vehicle malfunctions.

\subsubsection{Wind turbulence}

Wind turbulence affects an agent's ability to move in the environment.
We implement this phenomenon such that an agent's action will change to a neighboring action with probability $p$ (e.g., if the agent action is N, its neighboring actions are NE or NW). 

\subsubsection{Coverage area change}

This factor is intended to probe whether an agent trained on a smaller environment can map a larger environment, or vice versa. 

\section{Experiments}\label{sec:experiments}

We apply EMAC to the unknown environment coverage task, assessing its performance against the NRL, IQL, and IAC baselines. 
We discuss EMAC's advantages over these baselines and assess EMAC's robustness and scalability under a variety of environmental factors. 
Our last experiments study EMAC's performance using heterogeneous team compositions.

\subsection{Comparison to Baselines}\label{sec:baselines}

All methods are evaluated on an environment within an environment that has a 16$\times$16 grid, 3 agents, and an episode timeout of 100 steps. 
RL methods are trained for 15,000 episodes. During training of the RL methods, the average episode length of 40 independent trials are recorded every 100 time steps. Figure 4 plots the average episode length across training episodes for the IQL, IAC, and EMAC methods (NRL is a deterministic algorithm and does not require training). Table 1 compares the final results for all of the methods after training, averaged over 100 trials. 

\begin{table}[t]
\centering
 \begin{tabular}{ |c|c|c| } 
 \hline
 Algorithm & Completion time (steps) & Coverage  \\ 
 \hline\hline
 IQL & 40.86 & 0.99 \\ 
 \hline
 IAC & 18.43 & 1.00 \\
 \hline
 NRL & 19.80 & 1.00 \\
 \hline
 EMAC(ours) & 15.91 & 1.00 \\
 \hline
 \end{tabular}
 \caption{Comparison against baselines. EMAC outperforms traditional CPP approach NRL, as well as the two RL baselines, IQL and IAC.}
 \label{table:baselines}
\end{table}

\begin{figure}[t]
	\centering
   	\includegraphics{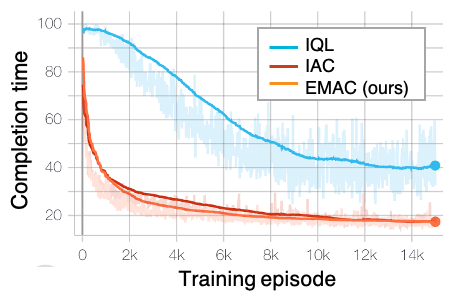}
	\caption{Average steps to complete the unknown environment coverage task over training episode. Compares the three RL methods (NRL is excluded since it is a deterministic algorithm).}
	\label{fig:baselines}
    \end{figure}

\begin{figure}[t]
	\centering
   	\includegraphics{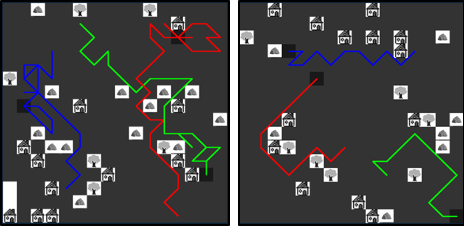}
	\caption{Visualization of EMAC flight paths. Left figure is EMAC at the beginning of training. Right figure is EMAC at the end of training – agent flight paths are much more direct and have little overlap with neighboring agents.}
	\label{fig:pathviz}
    \end{figure}

\begin{figure}[t]
	\centering
   	\includegraphics{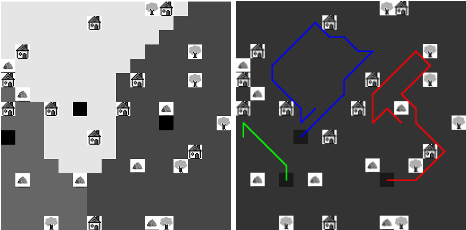}
	\caption{Visualization of NRL flight paths. Left figure is the Voronoi partitioning. Right figure shows the NRL flight paths. We observe that the green agent finishes early, but due to the restrictions of the partitioning, cannot help complete the red or blue agent's partitions, despite being near those sections.}
	\label{fig:nrl}
    \end{figure}

The results summarized in Table \ref{table:baselines} show that EMAC outperforms both the independent baselines IQL and IAC, as well as the NRL baseline. 
NRL has the advantage of a fully observable world for the environment decomposition step.
However, the resulting static partitioning limits the agents' ability to dynamically react to new information. 
For example, when visualizing NRL agent flight paths, we observed that agents would often finish their partitions at different times, but the assigned partitions limited the ability of early-finishing agents to assist late-finishing agents (Figure \ref{fig:nrl}). 
In contrast, EMAC agent flight paths demonstrate more flexible terrain coverage (Figure \ref{fig:pathviz}), suggesting that RL provides an advantage over traditional CPP algorithms in unknown environment contexts. 

Figure \ref{fig:baselines} demonstrates the performance of the RL-based methods: EMAC, IQL, and IAC. 
IQL and IAC training is unstable due to non-stationarity of the environment, even with the regularizing effects of multiple parallel worker threads.
Throughout training, IQL fails to converge to a good policy that quickly covers the grid. 
This failure likely occurs because of the non-stationarity nature of the algorithm in multi-agent settings.
We find that IAC oscillates around an acceptable policy after a certain number of iterations. 
However, it still underperforms EMAC, and takes many more episodes to reach a good solution. 
We hypothesize that this decrease in performance is due to the lack of a centralized critic, which is known to accelerate and stabilize training.

Figures \ref{fig:pathviz} and \ref{fig:nrl} show samples of EMAC and NRL agents executing in the environment, respectively.
Qualitatively, we see that EMAC agents learn cooperative behaviors based on the centralized critic sharing information among the actors. 
As training progresses, EMAC agent paths become less circuitous and the agents appear to coordinate region exploration based on the dynamic information they observe. 
In contrast, the NRL agents in Figure \ref{fig:nrl} are stuck with their original flight plans and are unable to adapt to changes in the environment.

\subsection{Scalability Experiments}\label{sec:scalability}

In these experiments, we more deeply probe EMAC's ability to scale to larger team sizes and larger environment grids. To test team size scalability, we train EMAC in a 22$\times$22 environment while  increasing the number of agents in the team from 2 to 8. We find that as the number of agents in the team increases, the number of steps it takes to complete the grid drops (Table \ref{table:agent_scalability}). These results indicate that EMAC performance is robust to large team sizes. Similarly, we find that EMAC's performance is robust to increasing environment sizes. In the environment scalability experiments, we train 3 agents in varying size environments from 16$\times$16 to 32$\times$32 (Table \ref{table:env_scalability}). As expected, we see a moderate increase in completion time as environment area increases. These results are in line with our basic expectations for the algorithm behavior, indicating that EMAC performance does not deteriorate under large input sizes.

The scalability of EMAC to larger team sizes and environment sizes can be partially attributed to the fixed-length observation representations used in the multi-agent environment coverage task. For example, the visitation history arrays are by nature invariant to the number of agents in the environment. Similarly, the pooled far-field visit history array and near-field visitation array remain a constant size even as the environment grows. This gives EMAC an advantage during learning. If the input representation were to grow with team or environment size, EMAC's models would also have to grow accordingly. Optimizing over a growing number of parameters is difficult and requires many more episodes to converge to a good solution. Keeping observation representations to a fixed length, regardless of team or environment size, allows EMAC to scale to large team and environment sizes.
    
\begin{table}[t]
\centering
 \begin{tabular}{ |c|c|c| } 
 \hline
 Number of agents & Completion time (steps)  \\ 
 \hline\hline
 2 & 60.95\\ 
 \hline
 4 & 31.85  \\
 \hline
 8 & 16.20 \\
 \hline
 \end{tabular}
 \caption{Agent scalability experiments. As expected, completion time decreases as number of agents increases, suggesting that EMAC is scalable to large team sizes with no performance degradation. }
 \label{table:agent_scalability}
\end{table}

\begin{table}[t]
\centering
 \begin{tabular}{ |c|c|c| } 
 \hline
Environment size & Completion time (steps)  \\ 
 \hline\hline
 16$\times$ 16 & 15.91 \\ 
 \hline
 20 $\times$ 20 & 30.44  \\
 \hline
24 $\times$ 24 & 50.76 \\
 \hline
 28 $\times$ 28 & 82.85 \\
 \hline
 32 $\times$ 32 & 147.30 \\
 \hline
 \end{tabular}
 \caption{Environment scalability experiments. As environment size increases, completion time increases in a measured manner, indicating that EMAC scales well to larger environments.}
 \label{table:env_scalability}
\end{table}

\subsection{Robustness Experiments}\label{sec:robustness}

Table \ref{table:robustness} summarizes the results of EMAC execution with different environmental factors, as described in \ref{sec:env_factors}. 
Each of the scenarios was tested using the EMAC policy actor trained in Section \ref{sec:baselines} \emph{without} environmental effects in the training process. 

These experiments show that EMAC-based agents handle several perturbed environmental factors despite never being exposed to these disturbances at training time. 
For example, EMAC displays surprising robustness to communication delays. 
We observe that a 1 step delay in communication results in a 7.2\% increase in episode length, and a 4 step delay results in a moderate 21.9\% increase. 
This robustness is likely due to EMAC agents' tendency to spread out in the environment. 
This means that information communicated by neighboring agents usually only affects an agent's far-field visit history, which is down-sampled before being fed into the actor network. 
Position changes therefore likely do not change an agent's observation until a few time steps have accumulated. Additionally, EMAC was able to handle wind turbulence probabilities of up to 20\% with only moderate increases in coverage area times. At higher wind turbulence probabilities, we found that EMAC exhibited more degraded performance, taking 77.2\% longer to cover the grid when wind turbulence probability reached 40\%.

For agent dropout and coverage area changes, performance degraded quickly. We found that increases in environment size led to much larger increases in coverage area times. For example, while the 32$\times$32 grid is four times as large as the baseline 16$\times$16 grid the agent team was trained on, evaluating on a 32$\times$32 grid resulted in completion times that were 14 times longer. Agent dropout also severely degraded performance, although it is hard to attribute whether that performance dip was a result of algorithmic issues or the fact that less agents at a given time were available to cover the area. 
Qualitatively, we observe that surviving agents move to cover areas that would have been covered by the dropped agent. 

\begin{table}[t]
\centering
 \begin{tabular}{ |m{3cm}|m{2cm}|m{2cm}| } 
 \hline
 \multicolumn{2}{|c|}{Environmental factor} & Completion time (steps) \\
 \hline\hline
 \multicolumn{2}{|c|}{Baseline} & 15.91 \\
 \hline
 \multirow{2}{3em}{Agent dropout} & 1 agent & 24.50 \\ 
 \cline{2-3}
 & 2 agents & 41.81 \\
 \hline
 \multirow{3}{3em}{Communication delay} & 1 step & 17.06 \\ 
 \cline{2-3}
 & 2 steps & 19.40 \\
 \cline{2-3}
 & 4 steps & 19.08 \\
 \hline
 \multirow{3}{3em}{Wind turbulence} & 10\% prob & 18.73 \\ 
 \cline{2-3}
 & 20\% prob & 20.01 \\
 \cline{2-3}
 & 40\% prob & 28.19 \\
 \hline
 \multirow{3}{3em}{Coverage area change} & 20$\times$20 & 36.42 \\ 
 \cline{2-3}
 & 24$\times$24 & 62.32 \\
 \cline{2-3}
 & 28$\times$28 & 132.25 \\
 \cline{2-3}
 & 32$\times$32 & 235.74 \\
 \hline
 \end{tabular}
 \caption{Robustness experiments. Overall, EMAC displayed a measure of robustness against external environmental factors not seen during training. However, at large extremes, EMAC performance does break down.}
 \label{table:robustness}
\end{table}

\subsection{Heterogeneous Team Experiments}\label{sec:heterogeneous}

We perform experiments showing EMAC’s performance on teams with heterogeneous sensor payloads. In these experiments, we construct teams of two different types of agents: (1) small agents with a 7$\times$7 field of view, and (2) large agents with a 9$\times$9 field of view.
They are placed them in a 20$\times$20 environment with four different team compositions. 
Results of the four different team compositions on the unknown environment coverage task are summarized in Table \ref{table:het}. 

We find that algorithm performance improves as we increase the proportion of large agents in the team, dropping from an average 32.72 steps to cover the grid to 22.79 steps. The intermediate heterogeneous compositions of Team 2 and Team 3 follow this trend. These results suggest that the EMAC fixed-length encoding layer enables good policy learning across agents with different sensing inputs. 

\begin{table}[t]
\centering
 \begin{tabular}{ |c|c|c| } 
 \hline
 Team \# & Team Composition & Completion time (steps)  \\ 
 \hline\hline
 1 & small, small, small & 32.72 \\ 
 \hline
 2 & small, small, large & 27.58  \\
 \hline
 3 & small, large, large & 24.28 \\
 \hline
 4 & large, large, large & 22.79 \\
 \hline
 \end{tabular}
 \caption{Heterogeneous experiments. As the number of larger agents (9$\times$9 FoV) increases, the number of training steps it takes to cover the grid decreases. EMAC is able to seamlessly handle both homogeneous (Team 1, Team 4) and heterogeneous (Team 2, Team 3).}
 \label{table:het}
\end{table}

\section{Conclusions}

In this paper, we presented EMAC, an algorithm for training decentralized policies in multi-agent settings that is extensible to homogeneous and heterogeneous team compositions. We assess the performance of our approach against a new multi-agent learning task, unknown environment coverage. This task has its roots in key government use cases such as disaster recovery and hazardous terrain mapping. In these scenarios, information about surrounding unknown areas would enhance the situational awareness of personnel on the field, facilitating decision-making in high-tempo environments.

Our results show that EMAC outperforms traditional multi-agent CPP techniques and independent RL baselines on the unknown environment coverage task. Unlike traditional CPP techniques that are limited by static, pre-computed agent flight paths, EMAC-enabled agents work in scenarios with unknown environmental elements such as dynamic obstacles, agent dropout, wind turbulence, and communication delays. 
Visualizations of agent flight paths demonstrate the flexible and cooperative behaviors that emerge from EMAC's trained policies. 
In addition, we demonstrate EMAC's scalability to large team sizes and environment sizes, and show EMAC's flexibility to various homogeneous and heterogeneous team compositions. 

In future work, we will investigate EMAC's extension to 3D environments and other government use cases that can be mapped to the unknown environment coverage task. EMAC provides flexibility and robustness to operate in unknown environments that would provide benefits to a number of key government use cases, such as disaster recovery, search and rescue efforts, and battlefield management. In these scenarios, an autonomous capability for hazardous terrain mapping could provide critical information about the surrounding area and enhance the situational awareness of personnel on the field.

\bibliography{references}
\bibliographystyle{aaai} 
\end{document}